\documentclass[iop]{emulateapj}

\usepackage{natbib}
\usepackage{gensymb}
\usepackage{hyperref}
\usepackage{graphicx}
\usepackage{multirow}
\usepackage{array}

\newcommand{\msun}{$\mbox{M}_{\odot}$}
\newcommand{\mstar}{$M_*$}

\newcommand{\secpoint}{\mbox{$''\mskip-7.6mu.\,$}}

\shorttitle{The MOSDEF Survey: [O~\textsc{iii}]$\lambda$4363 and direct-method oxygen abundance at $z=3.08$}
\shortauthors{Sanders et al.}

\begin{document}

\title{The MOSDEF Survey: Detection of [O~\textsc{iii}]$\lambda$4363 and the direct-method oxygen abundance of a star-forming galaxy at $\lowercase{z}=3.08$\altaffilmark{*}}
\altaffiltext{*}{Based on data obtained at the
W.M. Keck Observatory, which is operated as a scientific partnership among the California Institute of Technology, the
University of California, and NASA, and was made possible by the generous financial support of the W.M. Keck Foundation.
}

\author{Ryan L. Sanders\altaffilmark{1}}

\author{Alice E. Shapley\altaffilmark{1}}

\author{Mariska Kriek\altaffilmark{2}}

\author{Naveen A. Reddy\altaffilmark{3,4}}

\author{William R. Freeman\altaffilmark{3}}

\author{Alison L. Coil\altaffilmark{5}}

\author{Brian Siana\altaffilmark{3}}

\author{Bahram Mobasher\altaffilmark{3}}

\author{Irene Shivaei\altaffilmark{3}}

\author{Sedona H. Price\altaffilmark{2}}

\author{Laura de Groot\altaffilmark{6}}

\altaffiltext{1}{Department of Physics \& Astronomy, University of California, Los Angeles, 430 Portola Plaza, Los Angeles, CA 90095, USA}
\altaffiltext{2}{Astronomy Department, University of California, Berkeley, CA 94720, USA}
\altaffiltext{3}{Department of Physics \& Astronomy, University of California, Riverside, 900 University Avenue, Riverside, CA 92521, USA}
\altaffiltext{4}{Alfred P. Sloan Fellow}
\altaffiltext{5}{Center for Astrophysics and Space Sciences, University of California, San Diego, 9500 Gilman Dr., La Jolla, CA 92093-0424, USA}
\altaffiltext{6}{Department of Physics and Astronomy, Denison University, Granville, OH 43023, USA}

\email{email: rlsand@astro.ucla.edu}

\begin{abstract}
We present measurements of the electron-temperature based oxygen abundance for a highly star-forming galaxy
 at $z=3.08$, COSMOS-1908.  This is the highest redshift at which [O~\textsc{iii}]$\lambda$4363 has been detected,
 and the first time that this line has been measured at $z>2$.  We estimate an oxygen abundance of 12+log(O/H)$=8.00^{+0.13}_{-0.14}$.
  This galaxy is a low-mass ($10^{9.3}$~\msun), highly star-forming ($\sim50$~\msun~yr$^{-1}$) system that hosts a young stellar population ($\sim160$~Myr).
  We investigate the physical conditions of the ionized gas in COSMOS-1908 and find that this galaxy has a high
 ionization parameter, little nebular reddening ($E(B-V)_{\rm gas}<0.14$), and a high
 electron density ($n_e\sim500$~cm$^{-3}$).  We compare the ratios of strong oxygen, neon,
 and hydrogen lines to the direct-method oxygen abundance for COSMOS-1908 and additional
 star-forming galaxies at $z=0-1.8$ with [O~\textsc{iii}]$\lambda$4363 measurements, and
 show that galaxies at $z\sim1-3$ follow the same strong-line correlations as galaxies in the local universe.  This agreement
 suggests that the relationship between ionization parameter and O/H is similar for $z\sim0$ and high-redshift galaxies.
  These results imply that metallicity calibrations based on lines of oxygen, neon, and hydrogen
 do not strongly evolve with redshift and can reliably estimate abundances out to $z\sim3$, paving
 the way for robust measurements of the evolution of the mass-metallicity relation to high redshift.
\end{abstract}


\section{Introduction}\label{sec:intro}

The gas-phase metallicity
 of a galaxy is intimately connected to the processes governing galaxy formation and growth,
 namely the fueling and regulation of star formation.
  This connection is observed as the mass-metallicity relation (MZR), a correlation
 between stellar mass (\mstar) and gas-phase oxygen abundance for local star-forming galaxies
 \citep[e.g.,][]{tre04,and13}.  The MZR exists for high-redshift galaxies as well, and is observed
 to shift toward lower metallicity at fixed \mstar\ out to $z\sim3$ \citep[e.g.,][]{erb06,tro14,san15,ono16}.
  Constraining the evolution of the shape and normalization of the MZR with redshift provides
 insight into how the interplay among star formation, gas accretion, and feedback changes over cosmic history.
  This approach requires robust estimates of the gas-phase metallicity at all epochs.

The ratio of the flux of the auroral [O~\textsc{iii}]$\lambda$4363 line to that of
 [O~\textsc{iii}]$\lambda\lambda$4959,5007 is sensitive to the electron temperature of the ionized gas.
  Based on estimates of the electron temperature and density, the oxygen abundance can be inferred from
 ratios of strong oxygen lines (e.g., [O~\textsc{iii}]$\lambda\lambda$4959,5007
 and [O~\textsc{ii}]$\lambda$3727) to Balmer lines.
  Measurements of [O~\textsc{iii}]$\lambda$4363 have now been
 obtained for several hundred local H~\textsc{ii} regions and galaxies \citep[e.g.,][]{izo06,mar13}.
  However, the direct electron-temperature based method cannot be applied
 to the majority of local galaxies with spectroscopic observations because [O~\textsc{iii}]$\lambda$4363 is typically
 $\sim100$ times weaker than [O~\textsc{iii}]$\lambda$5007 at low metallicity, and much
 weaker still at solar and higher abundances.
  For this reason, calibrations to determine
 metallicity from strong optical emission-line ratios have been constructed based on observations
 of H~\textsc{ii} regions and galaxies with direct metallicity measurements
 \citep[e.g.,][]{pet04,pil05}.

Strong-line calibrations have been widely applied at both low and high redshifts, but it is
 uncertain whether these local calibrations can accurately predict the metallicities of
 high-redshift galaxies.  There is evidence that the physical conditions of the interstellar medium (ISM)
 in high-redshift galaxies differ from those observed locally.  In order to explain offsets
 between $z\sim0$ and high-redshift galaxies in diagnostic plots such as the
 [O~\textsc{iii}]/H$\beta$ vs. [N~\textsc{ii}]/H$\alpha$ diagram, it has been proposed that
 high-redshift galaxies may have higher ionization parameters \citep{kew15,cul16},
 harder ionizing stellar spectra \citep{ste14},
 higher density/ISM pressure \citep{kew13t,san16},
 and/or anomalous nitrogen abundance at fixed O/H \citep{mas14,sha15,san16}
 compared to $z\sim0$ galaxies.  Depending on which conditions evolve and the magnitude of
 that evolution, the relation between emission-line ratios and metallicity may
 change with redshift, potentially rendering current strong-line calibrations significantly biased at high redshifts.

Unbiased metallicity estimates for high-redshift galaxies based on the direct method are needed to
 evaluate the applicability of local metallicity calibrations at high redshift.  However,
 due to the weakness of [O~\textsc{iii}]$\lambda$4363 and the difficulties of observing in the near-infrared,
 this line has only been detected in 7 galaxies at $z>1$
 \citep{yua09,bra12b,chr12,sta13,jam14,mase14}, most of which are gravitationally lensed,
 and has not been detected at $z>1.9$.
  The small, inhomogeneous sample of direct-method
 metallicities at $z>1$ has made it difficult to assess the state of metallicity calibrations
 at high redshifts.

In this letter, we present observations of COSMOS-1908, an unlensed star-forming galaxy
 at $z=3.08$ with a detection of [O~\textsc{iii}]$\lambda$4363, observed as part of
 the MOSFIRE Deep Evolution Field (MOSDEF) survey \citep{kri15}.  We investigate the physical conditions
 of nebular gas in COSMOS-1908, considering multiple emission lines to evaluate the utility of
 locally-calibrated strong-line metallicity relations at $z\sim3$.  In Section~\ref{sec:obs}, we present
 details about the observations and data reduction.  We describe measurements of the spectral features
 in Section~\ref{sec:4363}.  In Section~\ref{sec:prop}, we derive galaxy and ionized gas properies of COSMOS-1908.
  Finally, we discuss the implications of our results in Section~\ref{sec:disc}.
  Throughout this paper, the term ``metallicity" refers to the gas-phase oxygen abundance (12+log(O/H)).
  We adopt a $\Lambda$-CDM cosmology with $H_0=70$~km~s$^{-1}$~Mpc$^{-1}$, $\Omega_m=0.3$, and $\Omega_{\Lambda}=0.7$.

\section{Observations and Reduction}\label{sec:obs}

We utilized data from the MOSDEF survey, described in detail
 in \citet{kri15}.  The data were obtained on 23 December 2012 and 1 April 2013 using the
 MOSFIRE spectrograph \citep{mcl12} on the 10~m Keck~I telescope.  COSMOS-1908 was observed
 in H and K bands on a MOSFIRE mask with 0\secpoint7 wide slits, resulting in spectral
resolutions of $\sim3650$ and $\sim3600$ in H and K, respectively.  Individual exposures
 were 120 seconds in H and 180 seconds in K.  The mask was observed for 72 minutes in H and
 144 minutes in K using an ABBA dither pattern with a 1\secpoint2 nod in December, and 64 minutes in H
 using an ABA'B' dither pattern with 1\secpoint5 and 1\secpoint2 outer and inner nods in April.  The
 total integration time was 136 minutes in H and 144 minutes in K.

The data were reduced using a custom IDL pipeline that produces a two-dimensional
 science and error spectrum for each slit on the mask that has been flatfielded, sky-subtracted, cleaned of cosmic rays,
 wavelength-calibrated, and flux-calibrated \citep{kri15}.
  One-dimensional science and error spectra were optimally extracted from the two-dimensional
 spectra.  Science spectra were corrected for slit losses on an individual basis using
 \textit{Hubble Space Telescope} F160W
 images and the average seeing for each filter.  Emission-line fluxes were measured by fitting
 Gaussian profiles to the one-dimensional science spectrum.  Uncertainties on line fluxes and all derived
 and measured quantities were estimated using a Monte Carlo technique, where the $1\sigma$ uncertainty bounds
 were taken to be the 16th and 84th percentile of the cumulative distribution function for each value.

In addition to our MOSFIRE observations,
 we utilized extensive multiwavelength photometric data that are available in the COSMOS field.
  COSMOS-1908 has coverage in 44 broad-, medium-, and intermediate-band filters from optical to
 infrared (rest-UV to rest-NIR), assembled by the 3D-HST team \citep{ske14,mom15}.

\section{The detection of auroral [O~\textsc{iii}]$\lambda$4363}\label{sec:4363}

\begin{figure*}
 \centering
 \includegraphics[width=\textwidth]{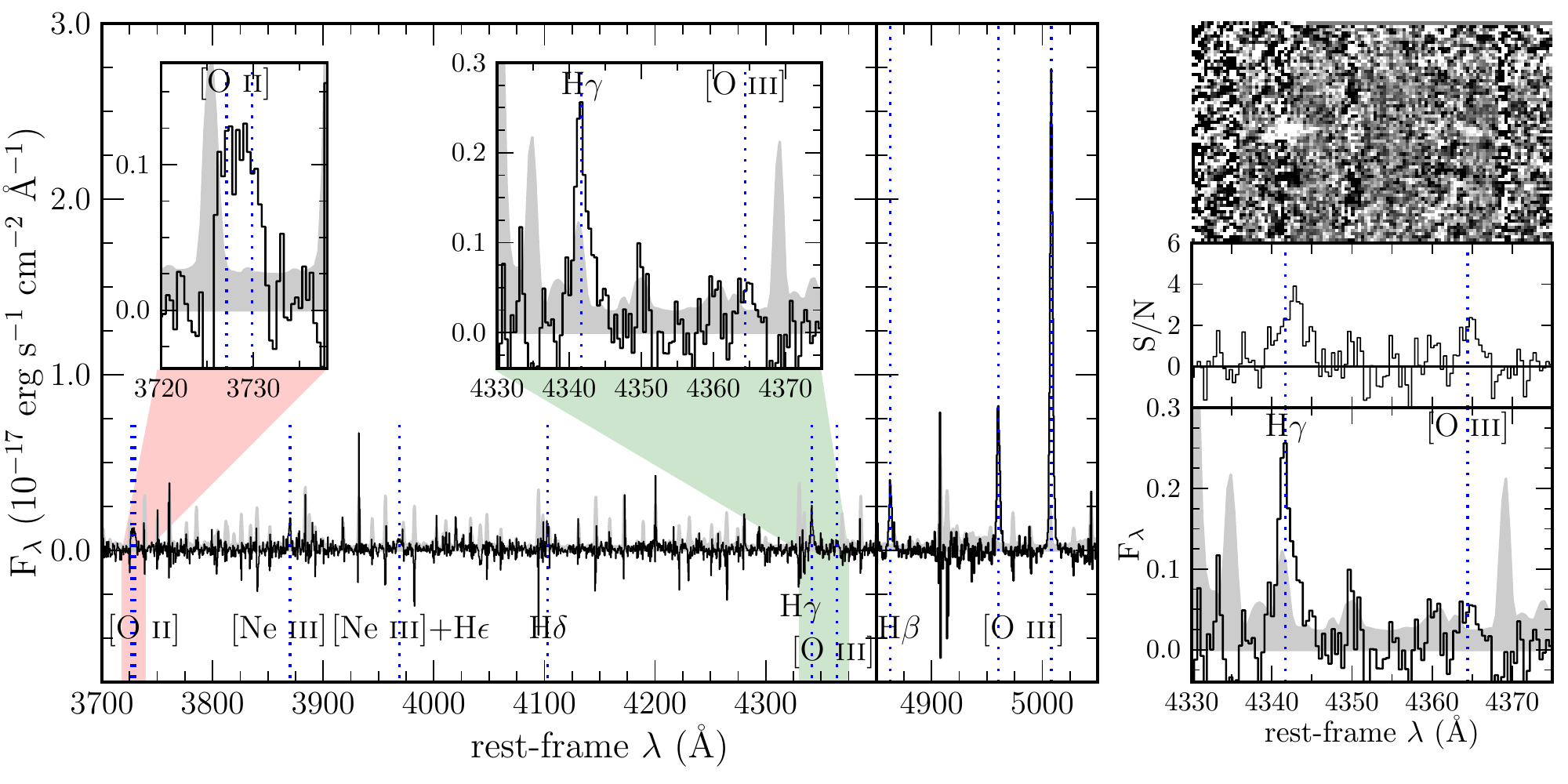}
 \caption{Spectrum of COSMOS-1908.  \textsc{Left}: H- and K-band spectra of COSMOS-1908.  The gray shaded region
 displays the magnitude of the error spectrum.  Emission lines
 are labeled, and the spectral regions around [O~\textsc{ii}] and [O~\textsc{iii}]$\lambda$4363 are highlighted.
  \textsc{Right}: Zoom-in of the wavelength region around H$\gamma$ and [O~\textsc{iii}]$\lambda$4363, showing the
 one-dimensional spectrum (bottom), signal-to-noise ratio spectrum (middle), and two-dimensional spectrum (top).
}\label{fig:spec}
\end{figure*}

Deep observations with MOSFIRE allow us to identify several emission lines in the
 H- and K-band spectra of COSMOS-1908, presented in Figure~\ref{fig:spec}.
  We measure a nebular redshift of $z=3.0768$ using the best-fit centroid of
 [O~\textsc{iii}]$\lambda$5007, the line with the highest signal-to-noise ratio (S/N).
  In addition to the strong rest-optical lines [O~\textsc{iii}]$\lambda\lambda4959,5007$, H$\beta$, and
 [O~\textsc{ii}]$\lambda\lambda$3726,3729, there are many weak emission lines detected in the H band.
  These include [Ne~\textsc{iii}]$\lambda$3869, [Ne~\textsc{iii}]$\lambda$3968 blended with
 H$\epsilon$, H$\delta$, H$\gamma$, and [O~\textsc{iii}]$\lambda$4363.  The observed emission-line fluxes and uncertainties
 are presented in Table~\ref{tab:table}.
  COSMOS-1908 displays a high level of excitation
 and ionization based on the strength of [Ne~\textsc{iii}] and the large [O~\textsc{iii}]$\lambda$5007 flux
 compared to that of H$\beta$ and [O~\textsc{ii}].

The auroral [O~\textsc{iii}]$\lambda$4363 line is detected with a formal signifiance of
 4.0$\sigma$.  The redshift of COSMOS-1908 fortuitously places [O~\textsc{iii}]$\lambda$4363 in a wavelength region that
 is free of skylines.
  The centroid of [O~\textsc{iii}]$\lambda$4363 implies a redshift ($z=3.0769$) that closely matches
 the one measured from stronger lines.  Additionally,
 emission can be seen at the expected location of [O~\textsc{iii}]$\lambda$4363 in the two-dimensional spectrum
 (Fig.~\ref{fig:spec}, right),
 and the S/N spectrum shows a coherent peak centered at the expected location of the line, where 7 consecutive
 pixels have S/N$>1$.
  We conclude that [O~\textsc{iii}]$\lambda$4363 is real and significantly detected, making COSMOS-1908 the
 highest-redshift galaxy for which [O~\textsc{iii}]$\lambda$4363 has been detected.

\begin{table}
 \centering
 \caption{Properties of COSMOS-1908.
 }\label{tab:table}
 \renewcommand{\arraystretch}{1.2}
 \begin{tabular}{ l c c }
   \hline\hline
   \multicolumn{3}{c}{Observed emission-line fluxes} \\
   \hline
   Line & Flux & Uncertainty \\[0pt]
        & {\scriptsize ($10^{-17}$ $\frac{\mbox{erg}}{\mbox{s cm}^2}$)} & {\scriptsize ($10^{-17}$ $\frac{\mbox{erg}}{\mbox{s cm}^2}$)} \\[2pt]
   \hline
   {[O~\textsc{ii}]$\lambda$3726} & 1.09 & 0.19 \\
   {[O~\textsc{ii}]$\lambda$3729} & 1.09 & 0.18 \\
   {[Ne~\textsc{iii}]$\lambda$3869} & 1.84 & 0.21 \\
   H$\delta$ & 1.49 & 0.32 \\
   H$\gamma$ & 2.21 & 0.34 \\
   {[O~\textsc{iii}]$\lambda$4363} & 0.56 & 0.14 \\
   H$\beta$ & 4.72 & 0.25 \\
   {[O~\textsc{iii}]$\lambda$4959} & 10.8 & 0.28 \\
   {[O~\textsc{iii}]$\lambda$5007} & 33.3 & 0.70 \\[2pt] 
   \hline\hline
   \multicolumn{3}{c}{Galaxy properties} \\[0pt]
   \hline
   Property && Value \\[0pt]
   \hline
   \multicolumn{2}{l}{Right Ascension (J2000)} & 10h 00m 23.36s \\[2pt]
   \multicolumn{2}{l}{Declination (J2000)} & $02\degree 11^{\prime} 55\secpoint9$ \\[2pt]
   $z$ && 3.0768 \\[0pt]
   log($M_*$/M$_{\odot}$) && $9.33^{+0.18}_{-0.17}$ \\[2pt]
   $E(B-V)_{\rm gas}$ && $0.0^{+0.14}_{-0.0}$ \\[2pt]
   SFR (M$_{\odot}$ yr$^{-1}$) && $49^{+30}_{-3}$ \\[2pt]
   sSFR (Gyr$^{-1}$) && $23^{+23}_{-6}$ \\[2pt]
   Area (kpc$^2$) && $4.4$ \\[2pt]
   \multicolumn{2}{l}{$\Sigma_{\rm{SFR}}$ (M$_{\odot}$ yr$^{-1}$ kpc$^-2$)} & $11^{+7}_{-1}$ \\[2pt]
   $n_e$ (cm$^{-3}$) && $520^{+600}_{-400}$ \\[2pt]
   {$T_e$([O~\textsc{iii}]) (K)} && $14000^{+1950}_{-1400}$ \\[2pt]
   {$T_e$([O~\textsc{ii}]) (K)} && $12800^{+1350}_{-1000}$ \\[2pt]
   12+log(O$^+$/H$^+$) && $6.87^{+0.17}_{-0.14}$ \\[2pt]
   12+log(O$^{++}$/H$^+$) && $7.96^{+0.13}_{-0.14}$ \\[2pt]
   12+log(O/H) && $8.00^{+0.13}_{-0.14}$ \\[2pt]
   \hline
 \end{tabular}
\end{table}

\section{Properties of COSMOS-1908}\label{sec:prop}

\subsection{The stellar content of COSMOS-1908}\label{sec:stellar}

\begin{figure*}
 \centering
 \includegraphics[width=0.725\textwidth]{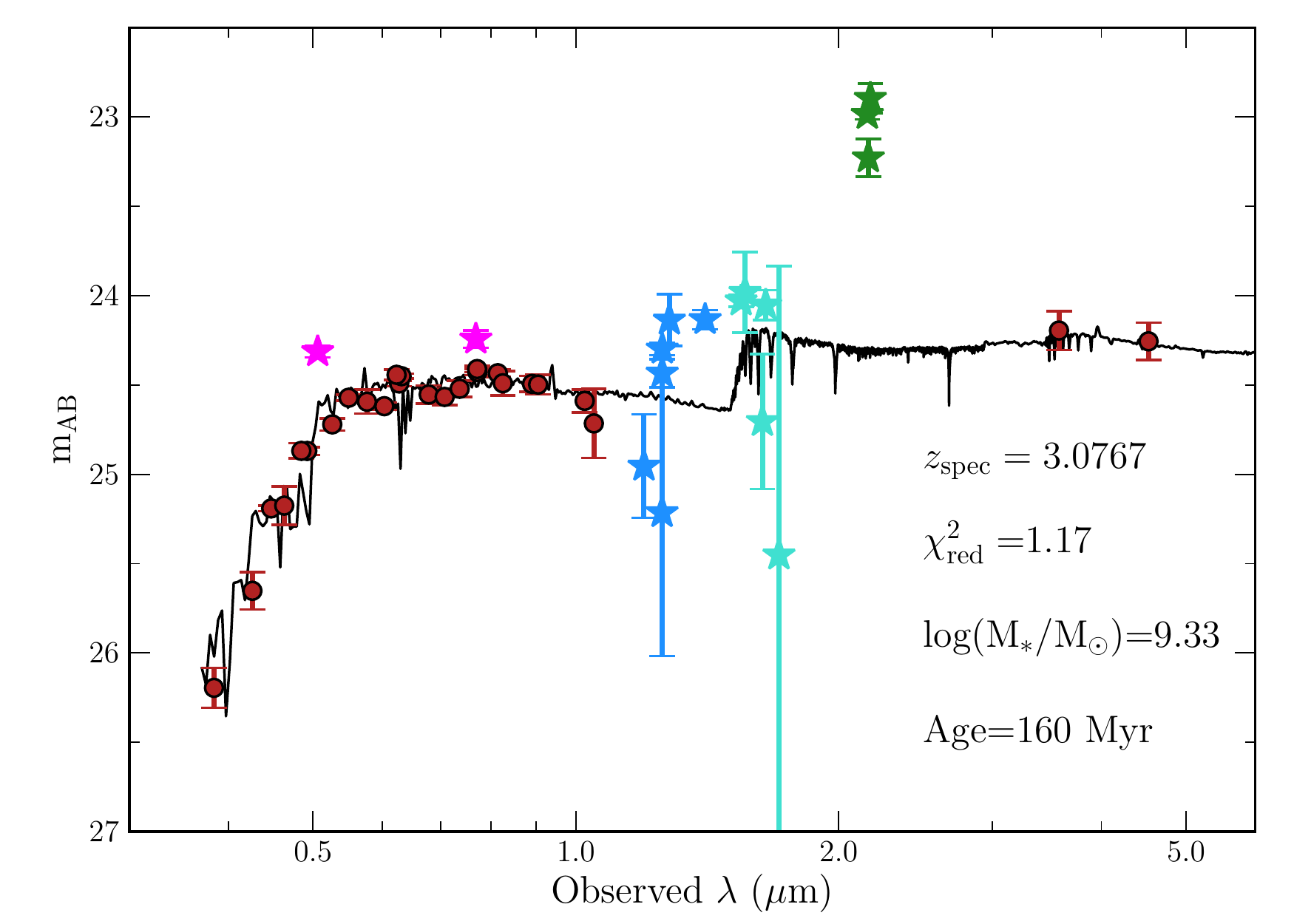}
 \includegraphics[width=0.265\textwidth]{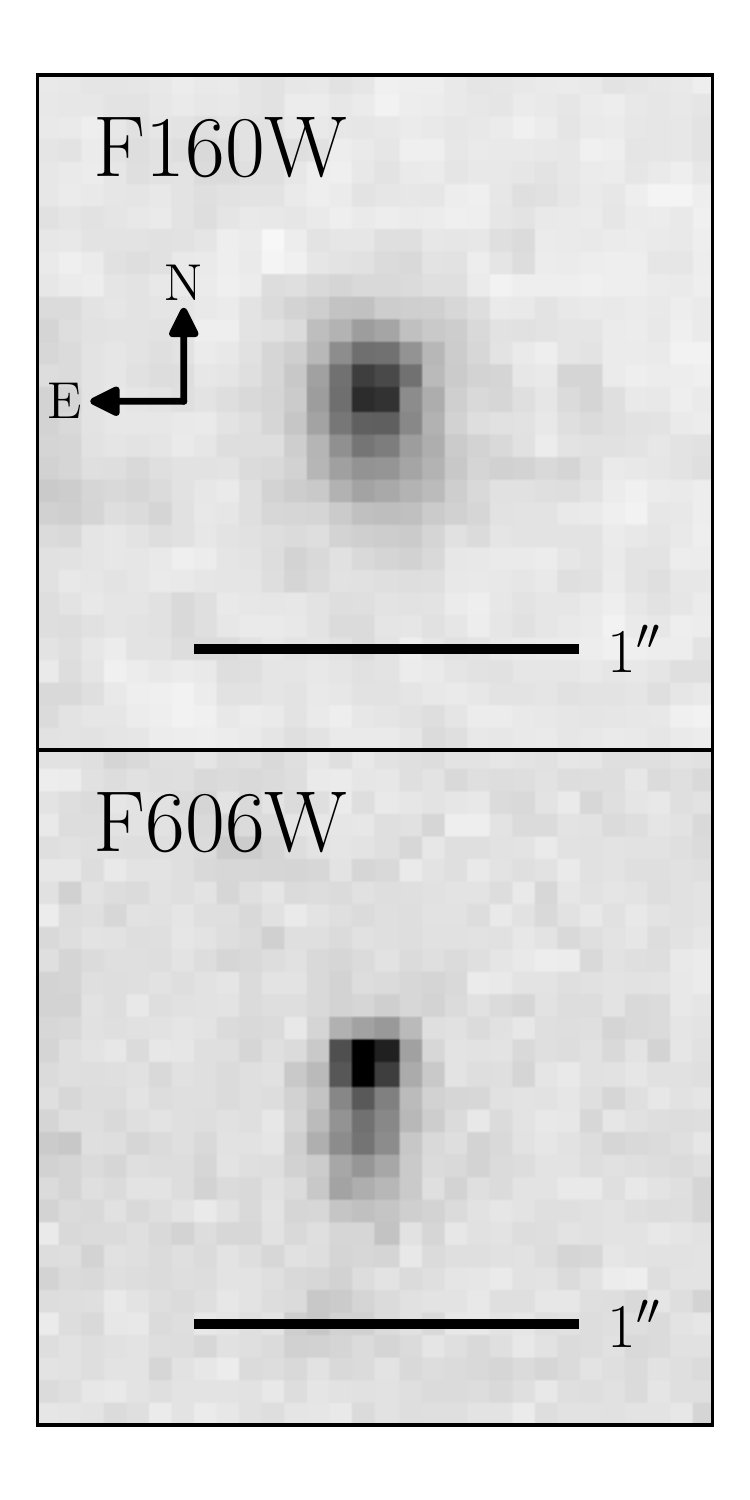}
 \caption{\textsc{Left}: Observed photometry and best-fit SED model for COSMOS-1908.  The red circles indicate
 photometry used for SED fitting.  Stars denote photometry excluded from the SED fitting process due to
 contamination.  Bands removed are: IA505 and IA767 (magenta), J (blue), H (cyan) and K (green).
  The 3D-HST photometric catalog includes multiple overlapping filters in the J, H, and K bands.
  \textsc{Right}: Postage stamp images of COSMOS-1908 in the F160W and F606W
 filters.  Images are 1\secpoint5 on a side, and the plate scale is 0\secpoint06/pixel.
}\label{fig:sed}
\end{figure*}

We estimate the stellar mass of COSMOS-1908
 via spectral energy distribution (SED) fitting.  The photometry is shown in Figure~\ref{fig:sed}.
  There is a clear excess in the K-band photometry due to
 emission-line flux from [O~\textsc{iii}]+H$\beta$ on top of the stellar continuum, as well
 as a probable excess in the H-band from weaker lines.
  For this reason, we exclude the H- and K-band photometric points when fitting the SED.
  The intermediate-band photometry at 5050~\AA\ and 7670~\AA\  were also excluded from the fit because of contributions
 from strong Ly$\alpha$ and C~\textsc{iii}]$\lambda$1909 emission, respectively.
  COSMOS-1908 additionally displays excess flux bluewards of the Balmer break in the J-band, which may be
 due to nebular continuum emission indicatave of a young stellar population \citep{nebcont}.
  Since our models are not tuned to fit nebular continuum emission, we exclude the J-band photometry as well.
  The remaining photometry was fit using the SED-fitting code FAST \citep{kri09}, with the \citet{con09} stellar
 population synthesis models, assuming a \citet{cha03} initial mass function, a delayed-$\tau$ star-formation history
 (${\rm SFR}\propto t e^{-t/\tau}$),
 and the \citet{cal00} attenuation curve.  We assume a stellar metallicity of 0.16~Z$_{\odot}$, the closest metallicity
 in the library of models to the measured gas-phase metallicity for COSMOS-1908 (see Section~\ref{sec:oh}).
  The photometry and best-fit SED model are shown in the left panel of Figure~\ref{fig:sed}.
  The best-fit age of the
 stellar population is $t=160^{+200}_{-110}$~Myr with an e-folding time of $\tau=10$~Gyr, indicating
 a rising star-formation history.  The extinction of the stellar continuum in the best-fit model is
 $E(B-V)_{\rm stars}=0.12^{+0.03}_{-0.05}$.
  The stellar mass is found to be log(\mstar/\msun)$=9.33^{+0.18}_{-0.17}$.
  Fitting including H- and K-band photometry that have been corrected for emission-line contamination
 using measured line fluxes changes $E(B-V)_{\rm stars}$ by $<0.05$, age by $\lesssim100$~Myr, and stellar mass
 by $<0.2$~dex.

The right panel of Figure~\ref{fig:sed} shows \textit{HST} postage stamp images of COSMOS-1908 in the
 F160W (rest-optical) and F606W (rest-UV) filters.  The galaxy displays a compact morphology that is slightly
 elongated towards the south.
  Using the F606W image, we measure a rest-UV area of 4.4~kpc$^2$ from the number of pixels
 that are 2$\sigma$ or more above the background noise.
  It is unlikely that COSMOS-1908 is dominated by ionization from an active galactic nucleus (AGN) due to its low stellar
 mass and the lack of a strong brightness peak in the central region.

\subsection{Nebular extinction and star-formation rate}\label{sec:sfr}

The reddening of nebular emission lines can significantly affect the inferred star-formation rate (SFR),
 line ratios, and derived metallicities.
  The amount of nebular extinction can be estimated using ratios of hydrogen Balmer-series lines.
  Balmer-series line fluxes are first corrected for stellar absorption using the best-fit SED model.
  Due to the strength of the emission lines, this correction is small, only 0.7\%, 2\%, and 3.5\% for H$\beta$, H$\gamma$,
 and H$\delta$, respectively.  Nebular reddening is estimated using Balmer-series line ratios, assuming
 intrinsic ratios of H$\gamma$/H$\beta=0.468$ and H$\delta$/H$\beta=0.259$ \citep{ost06} and the extinction curve
 of \citet{car89}.  The observed Balmer ratios of
 COSMOS-1908 are consistent with no reddening; $E(B-V)_{\rm gas}=0.0^{+0.14}_{-0.0}$, an upper limit that is
 consistent within the uncertainties with the stellar reddening.
  The SFR is estimated
 from H$\beta$ using the relation of \citet{ken98} converted to a \citet{cha03} IMF, assuming an intrinsic
 ratio of H$\alpha$/H$\beta$=2.847 \citep{ost06}.  COSMOS-1908 is vigorously forming stars, with
 SFR$=49^{+30}_{-3}$~\msun~yr$^{-1}$, specific SFR (sSFR; SFR/\mstar) of $23^{+23}_{-6}$~Gyr$^{-1}$,
 and SFR surface density ($\Sigma_{\rm{SFR}}$; SFR/Area) of $11^{+7}_{-1}$~\msun~yr$^{-1}$~kpc$^-2$.
  The galaxy properties of COSMOS-1908 are presented in Table~\ref{tab:table}.

\subsection{{\rm[O~\textsc{iii}]}$\lambda$5007 equivalent width}\label{sec:ew}

We measure a large [O~\textsc{iii}]$\lambda$5007 rest-frame equivalent width of
 EW([O~\textsc{iii}])=1600~\AA\ using the continuum level from the best-fit SED model.
  Objects with EW([O~\textsc{iii}]$)>1000$~\AA\ are extremely rare at $z\lesssim2.3$
 \citep{ate11}, but appear to be common at $z\sim6-7$ \citep{smi14}.
  Observing an object with such a high EW
 in the small survey volume of the MOSDEF $z>3$ observations suggests a significant increase
 in the occurrence rate of high-EW galaxies from $z\sim2$ to $z\sim3$.  Analysis of objects like COSMOS-1908
 can provide insight into the nature of $z\sim7$ star-forming galaxies with similar nebular EWs.
  COSMOS-1908 displays similar properties to the Lyman-continuum leaking galaxy Ion2 at $z=3.2$ \citep{deb16},
 including EW([O~\textsc{iii}]$)>1000$~\AA\ and [O~\textsc{iii}]$\lambda$5007/[O~\textsc{ii}]$\lambda3727>10$,
 and may be a good candidate for Lyman-continuum detection.

\subsection{Electron density}\label{sec:ne}

The electron density serves as a robust proxy for the hydrogen gas density in H~\textsc{ii} regions
 and can affect the strength of collisionally-excited lines.
  We calculate the electron density using the [O~\textsc{ii}]$\lambda$3729/$\lambda$3726 ratio (roughly unity) and
 the IRAF routine \textsc{nebular.temden} \citep{temden} with updated O~\textsc{ii} atomic data following
 \citet{san16}.  We assume an electron temperature of 12,800~K,
 the estimated $T_e$([O~\textsc{ii}]) for COSMOS-1908 (see Section~\ref{sec:oh}).
  We find the electron density to be $n_e=520^{+600}_{-400}$~cm$^{-3}$.  This high
 electron density compared to those of local star-forming galaxies and H~\textsc{ii} regions
 is in agreement with the order-of-magnitude
 increase in electron density from $z\sim0$ to $z\sim2.3$ observed in \citet{san16}.

\subsection{Electron temperature and oxygen abundance}\label{sec:oh}

We estimate the oxygen abundance following the prescription of \citet{izo06}.  The relative population of
 oxygen in ionization states higher than O$^{++}$ is assumed to be negligible, such that the
 total oxygen abundance is
\begin{equation}
\frac{\mbox{O}}{\mbox{H}}\approx\frac{\mbox{O}^+}{\mbox{H}^+}+\frac{\mbox{O}^{++}}{\mbox{H}^+}.
\end{equation}
The analytic equations of \citet{izo06} for ${\mbox{O}^+}/{\mbox{H}^+}$ and ${\mbox{O}^{++}}/{\mbox{H}^+}$
 require knowledge of the electron density, electron temperatures in the O$^+$ and O$^{++}$ zones, and
 dust-corrected emission-line fluxes for [O~\textsc{ii}], H$\beta$, and [O~\textsc{iii}].

We calculate $T_e$([O~\textsc{iii}]) using \textsc{nebular.temden} with updated O~\textsc{iii} collision
 strengths from \citet{stor14} and transition probabilities from the NIST MCHF database
 \citep{nistmchf}.
  In the density regime
 $n_e\lesssim1,000$~cm$^{-3}$, $T_e$ is insensitive to changes in density, so we do
 not iteratively solve for $n_e$ and $T_e$ simultaneously.  For COSMOS-1908, we find
 [O~\textsc{iii}]$\lambda\lambda$4959,5007/$\lambda4363=80.0^{+23.1}_{-20.0}$, which corresponds to an electron
 temperature in the O$^{++}$ zone of $T_e$([O~\textsc{iii}])=14,000$^{+1950}_{-1400}$~K.  We do not have
 wavelength coverage of the auroral [O~\textsc{ii}]$\lambda\lambda7320,7330$ lines to determine
 $T_e$([O~\textsc{ii}]) directly.
  Instead, we assume the linear $T_e$([O~\textsc{iii}])$-T_e$([O~\textsc{ii}]) relation of \citet{cam86}.
  We find an electron temperature in the O$^+$ zone of $T_e$([O~\textsc{ii}])=12,800$^{+1350}_{-1000}$~K.
  The total oxygen abundance of COSMOS-1908 is found to be 12+log(O/H)$=8.00^{+0.13}_{-0.14}$
 \citep[0.2~Z$_{\odot}$;][]{asp09}.
  The oxygen abundance, ionic abundances, electron temperatures, and density are listed in Table~\ref{tab:table}.

\section{Discussion}\label{sec:disc}

We investigate the evolution in the relationship between emission-line ratios and metallicity by
 comparing COSMOS-1908 to galaxies at lower redshifts with abundance determinations based on
 [O~\textsc{iii}]$\lambda$4363.  Recently, \citet{jon15} found that relations between direct-method oxygen abundance
 and ratios of neon, oxygen, and hydrogen emission lines do not evolve from $z=0-1$, using
 a sample of 32 star-forming galaxies at $z\sim0.8$ from the DEEP2 Galaxy Redshift Survey \citep{deep2}.
  We perform a similar comparison using galaxies at higher redshifts.
  The low-redshift comparison sample includes the $z\sim0.8$ galaxies from \citet{jon15} and 126 star-forming
 galaxies at $z\sim0$ from \citet{izo06} that have spectral coverage of [O~\textsc{ii}].
  We additionally compare to three galaxies at $z\sim1.5$
 \citep[][Mainali et al., in prep.]{chr12,jam14,sta13}.
  All galaxies in the comparison samples have detections of [O~\textsc{iii}]$\lambda$4363, and
 reddening corrections and oxygen abundances recalculated with the methods described in Section~\ref{sec:prop}.
  Uncertainties on emission-line ratios are calculated using a Monte Carlo technique, and include uncertainties
 in measurement and reddening correction.

\begin{figure*}
 \centering
 \includegraphics[width=\textwidth]{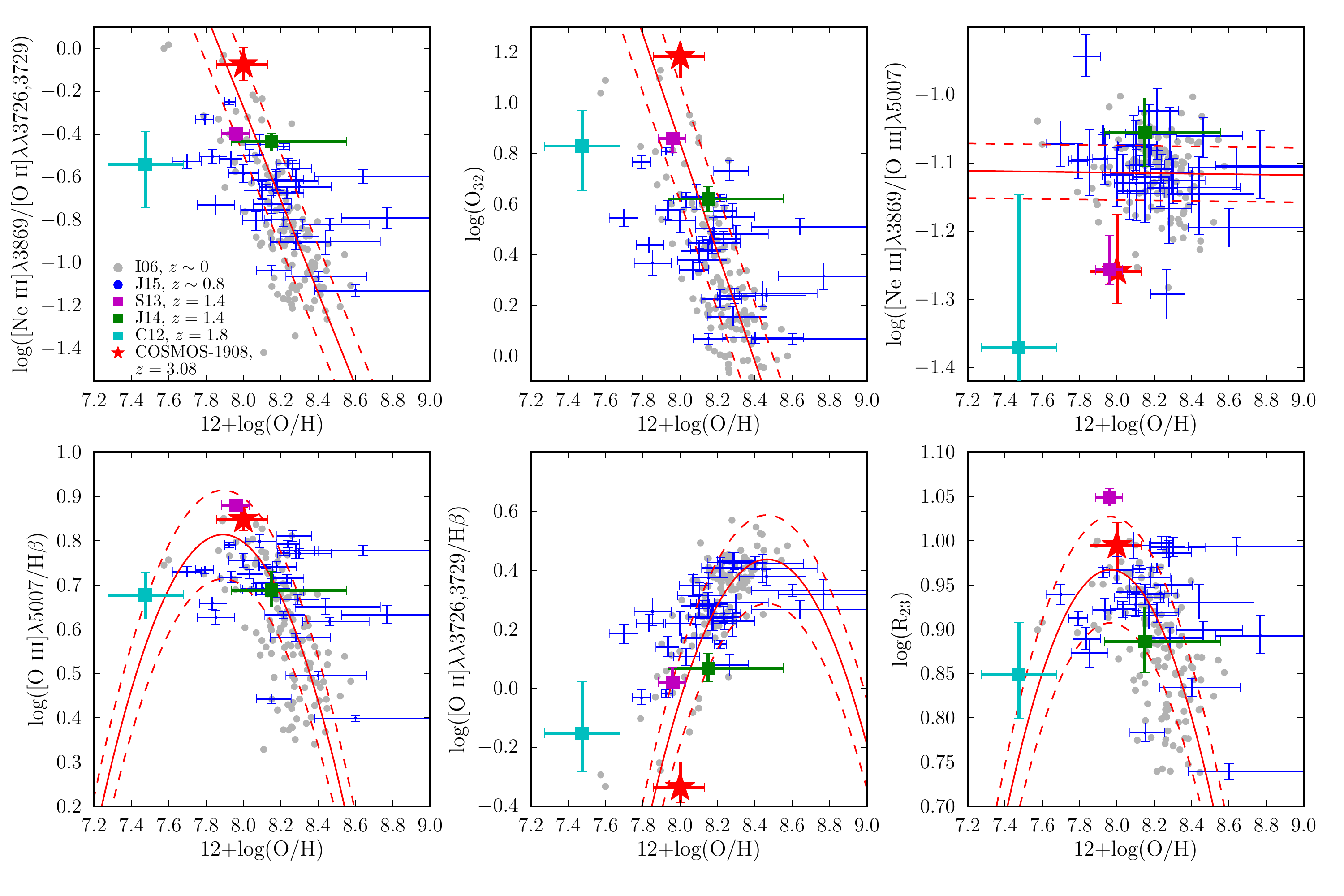}
 \caption{Emission-line ratios vs. direct-method oxygen abundance for COSMOS-1908 and lower-redshift comparison
 samples.  In each panel, the red star indicates COSMOS-1908.  Gray points show the $z\sim0$ sample from \citet[][I06]{izo06}, while
 blue errorbars show the positions of $z\sim0.8$ galaxies from \citet[][J15]{jon15}.  The red solid and dashed lines display
 the best-fit relations to the $z\sim0$ sample and 1$\sigma$ scatter \citep{jon15}, offset +0.045~dex in 12+log(O/H) to account for
 the different $T_e$([O~\textsc{iii}])$-T_e$([O~\textsc{ii}]) relation and O~\textsc{iii} atomic data used in this study.
  The magenta, green, and cyan squares show
 the $z=1.4$ galaxy from \citet[][S13; Mainali et al., in prep.]{sta13}, the $z\sim1.4$ galaxy from \citet[][J14]{jam14}, and
 the $z\sim1.8$ galaxy from \citet[][C12]{chr12}, respectively.
}\label{fig:lineratios}
\end{figure*}

The emission-line ratios and oxygen abundances are shown in Figure~\ref{fig:lineratios}.
  The line
 ratios displayed in each panel are sensitive to O/H, with the exception
 of [Ne~\textsc{iii}]/[O~\textsc{iii}], which should remain approximately constant with metallicity because it is a ratio
 of similar ionization states of $\alpha$-capture elements.
  In all 5 panels with metallicity-sensitive line ratios, COSMOS-1908 is consistent with the local best-fit
 relations within the uncertainties and instrinsic scatter.
  There are few galaxies
 in the low-redshift comparison samples that have the levels of low-metallicity and high-excitation that
 COSMOS-1908 displays, so this result relies on a small level of extrapolation of the $z\sim0$ relations.
  Of the four galaxies at $z>1$, COSMOS-1908 lies slightly towards the high-excitation side of the local relations,
 the \citet{sta13} and \citet{jam14} galaxies fall very nearly on each local relation, and the \citet{chr12} galaxy lies
 towards the low-excitation side.  Collectively, galaxies at $z>1$ do not show a systematic offset towards higher
 excitation at fixed metallicity, and the relation between these line ratios and O/H does not strongly evolve.

The close proximity of COSMOS-1908 to the local relations implies that evolution of the ionization parameter
 or ionizing spectrum at fixed O/H is small, if present.
  It has been suggested that high-redshift galaxies may have higher ionization
 parameters than local galaxies because of a scaling up of the radiation field due to more concentrated star formation
 \citep{kew15,cul16,kas16}.
  This scenario predicts that COSMOS-1908
 should have a much higher ionization parameter than local galaxies at similar O/H due to its extreme sSFR and
 $\Sigma_{\rm{SFR}}$, which would be observed as large offsets in
 [O~\textsc{iii}]/[O~\textsc{ii}], [O~\textsc{iii}]/H$\beta$, and [Ne~\textsc{iii}]/[O~\textsc{ii}] at fixed O/H.
  The models presented in \citet{cul16} predict that [O~\textsc{iii}]/H$\beta$ at fixed metallicity will
 be $>0.1$~dex larger at $z=3.1$ due to an increase in ionization parameter at fixed O/H.
  Such offsets are not observed in Figure~\ref{fig:lineratios}.

Our results instead suggest a scenario in which high-redshift and $z\sim0$ galaxies have similar
 ionization parameters at fixed metallicity, while high-redshift galaxies have higher ionization parameters
 than local galaxies at fixed \mstar\ due to the evolution of the MZR.  This scenario is consistent with our
 earlier findings in \citet{san16} using a sample of $\sim100$ star-forming galaxies at $z\sim2.3$ from the MOSDEF survey.
  The properties of COSMOS-1908 and the $z\sim1.5$ and $z\sim0.8$ comparison samples suggest that the same
 relationship between metallicity and ionization parameter holds out to $z\sim3$.
  Since changes in electron density minimally affect line ratios at subsolar metallicities \citep{san16},
  the agreement of
 the high- and low-redshift samples in Figure~\ref{fig:lineratios} implies that strong-line metallicity calibrations
 using only lines of oxygen, neon, and hydrogen can reliably predict abundances from $z\sim0$ to $z\sim3$.

Currently, these results are based upon a handful of galaxies at $z\sim1-3.5$, which is not sufficient to statistically
 constrain the behavior of the entire galaxy population at high redshifts.  More detections of [O~\textsc{iii}]$\lambda$4363
 for high-redshift galaxies are required to gain a complete understanding of the behavior of metallicity
 indicators at $z>1$.
  One way forward is selecting objects similar to COSMOS-1908 from photometric surveys by identifying objects
 with large excess flux in the photometric band covering [O~\textsc{iii}]$\lambda\lambda$4959,5007, suggestive of
 a large [O~\textsc{iii}] equivalent width and low metallicity.  Such objects are good candidates for [O~\textsc{iii}]$\lambda$4363
 detection in deep spectroscopic observations.
  Detections of weak features such as [O~\textsc{iii}]$\lambda$4363 for large samples at high redshift
 will be enabled by the next-generation near-infrared facilities such as the \textit{James Webb Space Telescope}
 and Thirty Meter Telescope.

\acknowledgements We would like to thank Tucker Jones, Dan Stark, and Ramesh Mainali for providing data used in this study.
  We acknowledge support from NSF AAG grants AST-1312780, 1312547, 1312764, and 1313171, and archival grant AR-13907,
 provided by NASA through the Space Telescope Science Institute.
  We wish to extend special thanks to those of Hawaiian ancestry on
 whose sacred mountain we are privileged to be guests. Without their generous hospitality, most
 of the observations presented herein would not have been possible.


\end{document}